\documentclass[%
 reprint, amsfonts,amssymb,amsmath,longbibliography,superscriptaddress]{revtex4-2}

\usepackage{graphicx}
\usepackage{dcolumn}
\usepackage{bm}
\usepackage[colorlinks=true,citecolor=blue,linkcolor=blue,urlcolor=blue]{hyperref}
\usepackage{color}
\usepackage{xcolor}
\usepackage[normalem]{ulem}

\begin{document}

\title{Comparative Raman study of Ruddlesden-Popper nickelates and the monolayer-trilayer polymorph}

\author{Vignesh~Sundaramurthy}
\affiliation{Max-Planck-Institute for Solid State Research, Heisenbergstra{\ss}e 1, 70569 Stuttgart, Germany}

\author{Abhi~Suthar}
\affiliation{Max-Planck-Institute for Solid State Research, Heisenbergstra{\ss}e 1, 70569 Stuttgart, Germany}

\author{Pascal~Puphal}
\affiliation{Max-Planck-Institute for Solid State Research, Heisenbergstra{\ss}e 1, 70569 Stuttgart, Germany}

\author{Congcong~Le}
\affiliation{Hefei National Laboratory, Hefei 230088, China}

\author{Yuhao~Gu}
\affiliation{School of Mathematics and Physics, University of Science and Technology Beijing, Beijing 100083, China}

\author{Hasan~Yilmaz}
\affiliation{Max-Planck-Institute for Solid State Research, Heisenbergstra{\ss}e 1, 70569 Stuttgart, Germany}
\affiliation{University of Stuttgart, Institute for Materials Science, Materials Synthesis Group, Heisenbergstraße 3, 70569 Stuttgart, Germany}

\author{Pablo~Sosa-Lizama}
\affiliation{Max-Planck-Institute for Solid State Research, Heisenbergstra{\ss}e 1, 70569 Stuttgart, Germany}

\author{Peter~A.~van~Aken}
\affiliation{Max-Planck-Institute for Solid State Research, Heisenbergstra{\ss}e 1, 70569 Stuttgart, Germany}

\author{Y.~Eren~Suyolcu}
\affiliation{Max-Planck-Institute for Solid State Research, Heisenbergstra{\ss}e 1, 70569 Stuttgart, Germany}

\author{Masahiko~Isobe}
\affiliation{Max-Planck-Institute for Solid State Research, Heisenbergstra{\ss}e 1, 70569 Stuttgart, Germany}

\author{Andreas~P.~Schnyder}
\affiliation{Max-Planck-Institute for Solid State Research, Heisenbergstra{\ss}e 1, 70569 Stuttgart, Germany}

\author{Xianxin~Wu}
\affiliation{Institute of Theoretical Physics, Chinese Academy of Sciences, Beijing 100190, China}

\author{Matteo~Minola}
\affiliation{Max-Planck-Institute for Solid State Research, Heisenbergstra{\ss}e 1, 70569 Stuttgart, Germany}

\author{Bernhard~Keimer}
\affiliation{Max-Planck-Institute for Solid State Research, Heisenbergstra{\ss}e 1, 70569 Stuttgart, Germany}

\author{Matthias~Hepting}
\email[]{hepting@fkf.mpg.de}
\affiliation{Max-Planck-Institute for Solid State Research, Heisenbergstra{\ss}e 1, 70569 Stuttgart, Germany}

\date{\today}

\begin{abstract}
Ruddlesden–Popper (RP) nickelates have attracted intense interest following the discovery of superconductivity in several members of the series, including bilayer (BL) La$_3$Ni$_2$O$_7$, trilayer (TL) La$_4$Ni$_3$O$_{10}$, and structural polymorphs composed of monolayer–bilayer or monolayer–trilayer (ML–TL) units. However, an inherent propensity of the RP series to form intergrown phases during single-crystal synthesis, together with spatial variations in oxygen stoichiometry, has complicated the determination of their intrinsic material properties. As a consequence, conflicting reports have emerged on both their electronic phase transitions and lattice dynamics. In this work, we perform a comparative study of the phononic and electronic Raman responses of high-quality ML–TL single crystals and contrast them with those of the other RP nickelates, using samples with optimized oxygen content. We establish several Raman spectral features that enable unambiguous phase identification across the series. Moreover, we uncover characteristics in the phononic and electronic Raman response of ML–TL that are not reflected in the pure ML and TL compounds. We attribute these differences to a distinctive electronic structure arising from self-doping and confinement effects induced by the ML unit within the ML–TL lattice architecture.
\end{abstract}

\maketitle

\section{Introduction}

The search for high-$T_c$ unconventional superconductivity beyond the layered cuprates has long been guided by the idea that materials sharing key structural and electronic configurations with cuprates might host similar physics \cite{Norman2016}. Shortly after the discovery of superconductivity in hole-doped cuprates, nickelates with composition $RE_2$NiO$_4$ ($RE$ = rare-earth ion) were intensively investigated as natural candidates \cite{Acrivos1994}. This interest was motivated by the chemical proximity of Ni and Cu and by the observation of charge and spin stripe order \cite{Sachan1995}, reminiscent of the stripe phases in hole-doped La$_2$CuO$_4$ \cite{Tranquada2020}, where they precede the onset of superconductivity \cite{Keimer2015,ScalapinoRevModPhys2012}. Despite extensive experimental and theoretical efforts, however, superconductivity has never been conclusively established in nickelates composed of single layers of NiO$_6$ octahedra \cite{Ji2024}, such as hole-doped $RE_{2-x}$Sr$_x$NiO$_4$ \cite{Uchida2011}. Superconductivity has likewise not been observed in nickelate heterostructures with ultra-thin perovskite layers \cite{Middey2016} that were predicted to realize a cuprate-like electronic structure \cite{Chaloupka2008,Hansmann2009}.

A critical advance came with the discovery of superconductivity in thin films of hole-doped infinite-layer (IL) nickelates \cite{LiNature2019,PuphalReview2025,Wang2024review}. These materials exhibit square-planar NiO$_2$ layers without apical oxygen stacked along the $c$ axis direction, a structural motif that had been predicted by early theoretical work to be favorable for cuprate-like electronic states \cite{AnisimovPRB1999}. Subsequent extensions to related IL-derived systems, including the quintuple-layer compound Nd$_6$Ni$_5$O$_{12}$ \cite{PanNatMater2022}, demonstrated that superconductivity is not restricted to a unique structural realization, but might instead emerge across a broader nickelate materials space. Indeed, superconductivity was subsequently observed in single crystals of the bilayer (BL) nickelate La$_3$Ni$_2$O$_7$, which contains apical oxygen, with transition temperatures exceeding 80~K under high pressure \cite{Sun2023}. Superconductivity was also reported in epitaxial thin films of the same compound, without the need for external pressure application \cite{KoNature2025, Zhou2025F}. The remarkably high $T_c$ in BL single crystals surpasses the boiling point of liquid nitrogen and therefore represents a major milestone in the search for new families of high-$T_c$ superconductors \cite{Hepting2023}. 

Both the BL compound and the earlier investigated monolayer (ML) nickelates belong to the Ruddlesden--Popper (RP) series \cite{Greenblatt1997}, described by the empirical formula $RE_{n+1}$Ni$_n$O$_{3n+1}$, which consists of $n$ perovskite-like NiO$_6$ layers separated by rock-salt $RE$O blocks. Following the BL compound, superconductivity was also reported in the trilayer (TL) nickelate La$_4$Ni$_3$O$_{10}$, albeit with a lower $T_c$ of 30 - 40 K \cite{ZhuNature2024, ZhangPRX2025}. This trend contrasts with the layer dependence observed in cuprates, where $T_c$ typically increases from single-layer to multilayer systems \cite{Keimer2015,Iyo2007}, and points to a distinct and nontrivial interplay between crystal structure and electronic properties in RP nickelates \cite{Wang2024b, ZhuNature2024, LuPRB2025, ZhangPRX2025, HuangPRB2024, Khaliullin2025, Bejas2025, SakakibaraPRB2024}.

Nonetheless, the superconducting phases of BL and TL nickelates emerge in close proximity to charge density wave (CDW) and spin density wave (SDW) instabilities, reminiscent of the behavior observed in cuprate and iron-based superconductors \cite{Keimer2015,ScalapinoRevModPhys2012,FernandesNature2022}. In TL nickelates, intertwined CDW and SDW order condense below approximately 140~K \cite{ZhangNatComm2020}, whereas in BL nickelates the two orders are disjoint, with $T_{\mathrm{SDW}} \approx 150$~K and $T_{\mathrm{CDW}} \approx 110$~K \cite{Liu2022}. These density-wave (DW) instabilities have been investigated using a variety of spectroscopic techniques, including angle-resolved photoemission spectroscopy (ARPES), muon spin rotation/relaxation ($\mu$SR), and Raman scattering \cite{Li2017,Liu_2024,Li2025SDW,Xu2025,Chen2024uSR, Khasanov2025,Kakoi2024, HeRSP2025, DeswalAPL2025, ramakrishnanarXiv2025, DeswalarXiv2025, GimPRL2025, Zhang327HighP2025}. Specifically, recent polarized Raman measurements have revealed a multiorbital character of the DW instability, involving both Ni $3d_{x^2-y^2}$ and $3d_{z^2}$ states \cite{SutharVS2025}.

Beyond the conventional RP phases with uniform stacking of NiO$_6$ octahedral units, a tendency to form unconventional polymorphic structures during bulk single-crystal growth has been reported, yielding variants with alternating monolayer--bilayer (ML--BL) \cite{LiPRM2024,ShiNature2025} or monolayer--trilayer (ML--TL) stacking sequences \cite{PuphalPRL2024,ChenACS2024,Wang13132024}. These materials are also referred to as the ``1212" and ``1313" phases, respectively. The ML--BL variant exhibits a $T_c$ of approximately 60~K \cite{ShiNature2025}, whereas for the ML--TL polymorph reported $T_c$ values range between 4 - 80~K \cite{PuphalPRL2024,HuangarXiv2025}. Whether specific polymorphic arrangements of RP building blocks favor or suppress DW correlations, or support a decoupling of spin and charge degrees of freedom, remains under debate as the reported trends vary substantially and are likely influenced by differences in oxygen stoichiometry among samples \cite{LiPRM2024,ShiNature2025,PuphalPRL2024,ChenACS2024,Auyeung2025,Wang13132024,HuangarXiv2025}. Nevertheless, these polymorphs offer an additional avenue for probing the layer dependence of correlated electron phenomena in nickelates. At the same time, they introduce an added level of complexity beyond the already multiband and multiorbital nature of conventional RP nickelates \cite{LuoNature2024,LuPRL2024,WangCPL2024,RyeePRL2024,JiangPRL2025}, since multiple structural building blocks contribute simultaneously to their low-energy electronic structure.

In particular, ARPES measurements on the ML--TL polymorph revealed the presence of the additional $\varepsilon$ pocket at the Fermi-surface, exhibiting Ni $3d_{z^2}$ orbital character \cite{AbadiPRL2025}, which is absent in pure ML, BL, and TL phases \cite{UchidaPRB2011,Li2017,Yang2024ARPES}. Furthermore, the quantum confinement of the TL between the ML blocks along with possible charge transfer was proposed to alter correlation strength and Ni valences \cite{LaBollitaPRB2024,ZhangPRB2024,LiuJPCC2025,Ouyang2025,Samanta2025}, possibly giving rise to distinct Fermi-surface features, such as Fermi arcs \cite{LechermannPRM2024}. It therefore remains an open and central question how these emergent electronic features in the polymorphs influence the DW instabilities and superconductivity, and more broadly how lattice degrees of freedom are intertwined with the low-energy electronic structure. 

In this work, we use polarized Raman scattering to investigate the phononic and electronic Raman response of the ML-TL polymorph, in direct comparison to the Raman response of other relevant members of the RP nickelates series. We observe unique Raman spectral features for each material, serving as a spectral fingerprint to identify the phase. Notably, the TL phase and the ML-TL polymorph exhibit a strikingly similar phonon spectrum at room temperature, although with subtle distinctions. At lower temperatures, dramatically different spectral features unique to each structural motif emerge, demonstrating the significance of Raman spectroscopy in the study of polymorph structures. In electronic Raman scattering, we observe emergent features attributable to DW formation, with both distinctions and similarities compared to those reported for the pure BL and TL compounds.

\section{Materials and Methods}

Single crystals of the RP series La$_{n+1}$Ni$_n$O$_{3n+1}$ were grown using a high-pressure high-temperature optical floating-zone (OFZ) furnace (Model HKZ, SciDre GmbH, Dresden). Details about sample synthesis and characterization are provided in the Supplemental Material. 

The Raman spectra were obtained with a HORIBA Jobin Yvon LabRAM HR800 spectrometer in backscattering geometry, using a He-Ne laser (wavelength 632.8 nm) focused on the sample with a x50 magnification ultra-long working distance objective. The laser power on the sample was less than 1~mW. The indicated temperatures were determined by Stokes/anti-Stokes intensity analysis. All plotted Raman spectra were divided by the Bose-Einstein factor. For low-temperature measurements, a Konti Micro helium-flow cryostat from CRYOVAC was used.

 Phonon frequency calculations were performed using density functional theory (DFT) as implemented in the Vienna \textit{ab-initio} simulation package (VASP) code~\cite{kresse1996}. The Perdew–Burke–Ernzerhof (PBE)~\cite{perdew1996} exchange–correlation functional and the projector-augmented-wave (PAW)~\cite{Joubert1999} approach were used. Lattice constants and internal atomic positions were relaxed using generalized gradient approximation (GGA), with a plane-wave cutoff energy of 600 eV. Forces were minimized to less than 0.001 $\mathrm{eV}$/{\AA} in the relaxation.  The calculations were performed for the tetragonal $P4/mmm$ unit cell. This symmetry is adopted by the ML-TL polymorph under pressure \cite{PuphalPRL2024} when the octahedral tilts characterizing the larger $Fmmm$ unit cell vanish. The relaxed structural parameters were $a,b$ = 3.84~{\AA} and $c$ = 20.24~{\AA}, obtained from the $P4/mmm$ structure under 1.5 GPa. In the phonon calculations, the cutoff energy was set to 400 eV for expanding the wave functions into plane-wave basis, and the number of $k$ points was set to $3\times4\times4$ for a $1\times2\times2$ supercell. The real-space force constants of the supercells were calculated using density-functional perturbation theory (DFPT)~\cite{Baroni1987} and the phonon frequencies were calculated from the force constants using the PHONOPY code~\cite{Togo2008,TOGO20151}.

\section{Results}

\subsection{Phase identification}

\begin{figure}[tb]
    \centering
    \includegraphics[width=\linewidth]{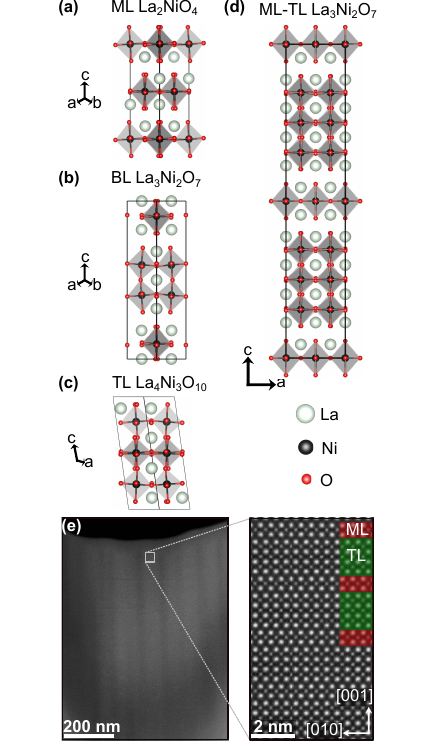}
    \caption{(a-d) Schematics of the crystal structures of (a) ML La$_2$NiO$_4$, (b) BL La$_3$Ni$_2$O$_7$, (c) TL La$_4$Ni$_3$O$_{10}$, and (d) ML-TL La$_3$Ni$_2$O$_7$, with $Bmab$, $Cmcm$, $P2_1/a$, and $Fmmm$ unit cells, respectively. (e) STEM-HAADF images of a ML-TL single crystal. The low-resolution image (left) demonstrates  absence of secondary phases on large length scales. The contrast between bright and dark vertical stripes corresponds to minor topographic differences due to curtaining effects caused by ion milling during specimen preparation. The atomic-resolution image (right) resolves the stacking sequence of ML (red) and TL (green) units. The [001] and [010] directions of the $Fmmm$ unit cell are indicated. }
    \label{fig:Lattice}
\end{figure}

\begin{figure}[tb]
    \centering
    \includegraphics[width=\linewidth]{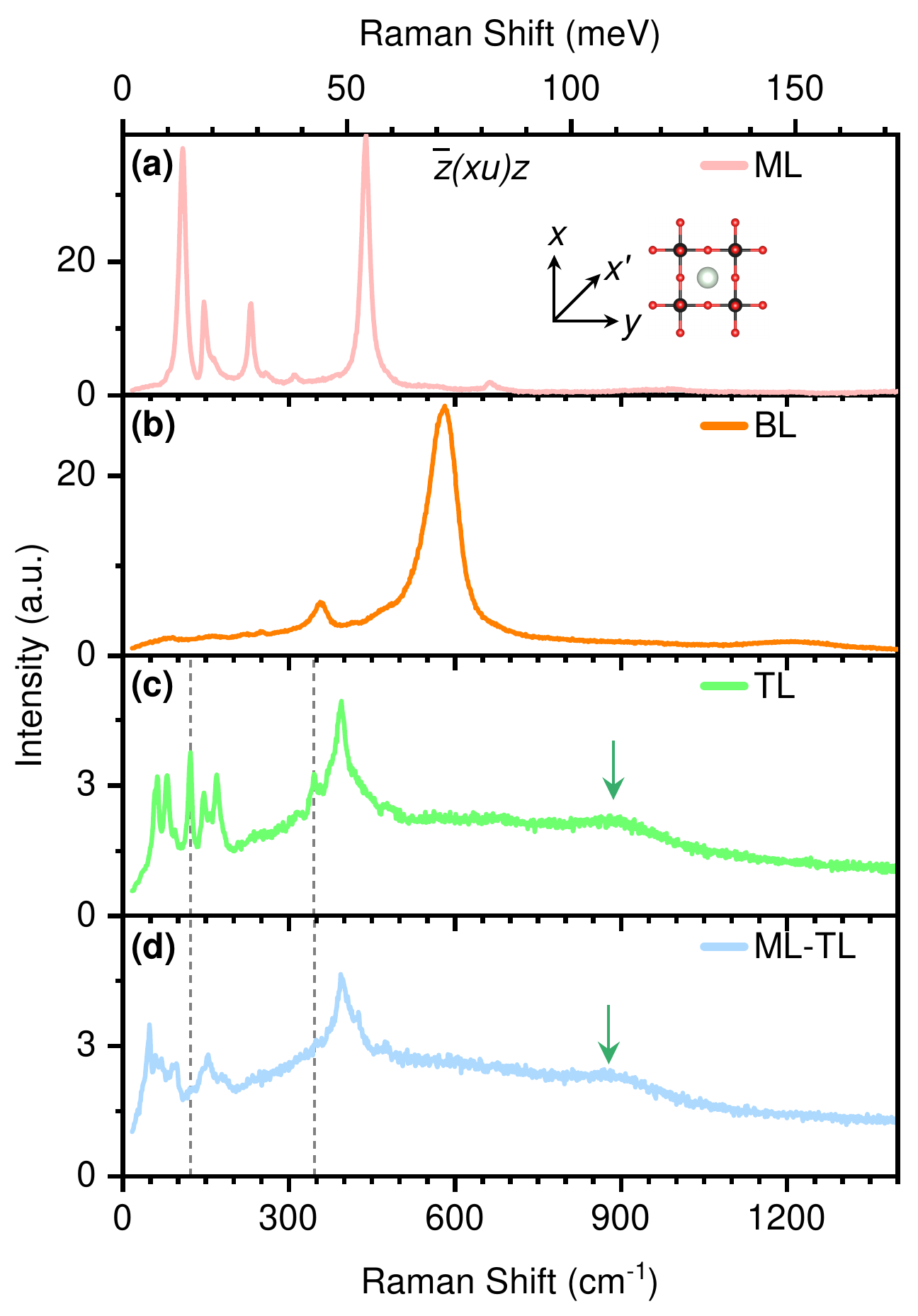}
    \caption{(a-d) Raman spectra of the (a) ML, (b) BL, (c) TL, and (d) ML-TL phases acquired at room temperature in $\bar z(xu)z$ scattering configuration (without analyzer). The inset in (a) shows the top-view of a Ni-O plaquette in the \textit{ab}-plane of layered nickelates, with octahedral tilts omitted for simplicity. The $x$ and $y$ directions of the laboratory frame of the Raman experiment point along the Ni-O-Ni bond directions. The $x'$ direction is also indicated. Vertical dashed lines in (c) and (d) indicate the position of characteristic phonon modes of the TL phase that are absent in ML-TL. Green arrows mark a broad electronic Raman feature present in both TL and ML-TL.
    }
    \label{fig:RoomT_Raman}
\end{figure}

Figures~\ref{fig:Lattice}(a-c) provide an overview of the lattice architectures of the ML La$_2$NiO$_4$, BL La$_3$Ni$_2$O$_7$, and TL La$_4$Ni$_3$O$_{10}$ RP nickelates, along with the ML-TL La$_3$Ni$_2$O$_7$ polymorph in Fig.~\ref{fig:Lattice}(d). A wide-area survey using high-angle annular dark-field (HAADF) scanning transmission electron microscopy (STEM) on an ML–TL single crystal used for the Raman measurements demonstrates high crystalline quality and the absence of secondary phases on macroscopic length scales (Fig.~\ref{fig:Lattice}(e), left panel). Atomic-resolution HAADF imaging from a representative area resolves the characteristic alternating stacking of ML and TL structural units (Fig.~\ref{fig:Lattice}(e), right panel). Note that phase-pure ML–TL single crystals grown by the OFZ technique reach lateral dimensions of several tens to a few hundred micrometres \cite{PuphalPRL2024,ChenACS2024,Auyeung2025}, whereas larger crystals typically contain mixed-phase regions. The spatial resolution of micro-Raman spectroscopy (see Materials and Methods) therefore enables measurements on small single-phase crystals as well as the selective probing of individual phases in mixed RP nickelate crystals. 

Figure~\ref{fig:RoomT_Raman} contrasts the distinct Raman spectra of the ML, BL, TL, and ML-TL phases. The scattering geometry including the photon polarizations are given in Porto's notation, with $u$ indicating that no analyzer was inserted in the scattered light path, \textit{i.e.}, $\bar z(xu)z$ contains both $\bar z(xx)z$ and $\bar z(xy)z$ scattering contributions. For simplicity, the labels $\bar z$ and $z$, denoting the incident and scattered photon directions in back-scattering geometry, respectively, are omitted in the following. The $x$ and $y$ directions are aligned parallel to the Ni-O-Ni bond directions within the reference frame of the square-planar NiO$_2$ planes. The $x^\prime$ and $y^\prime$ directions are aligned 45$^\circ$ to the Ni-O-Ni bonds.

\begin{figure*}[tb]
    \centering
    \includegraphics[width=\linewidth]{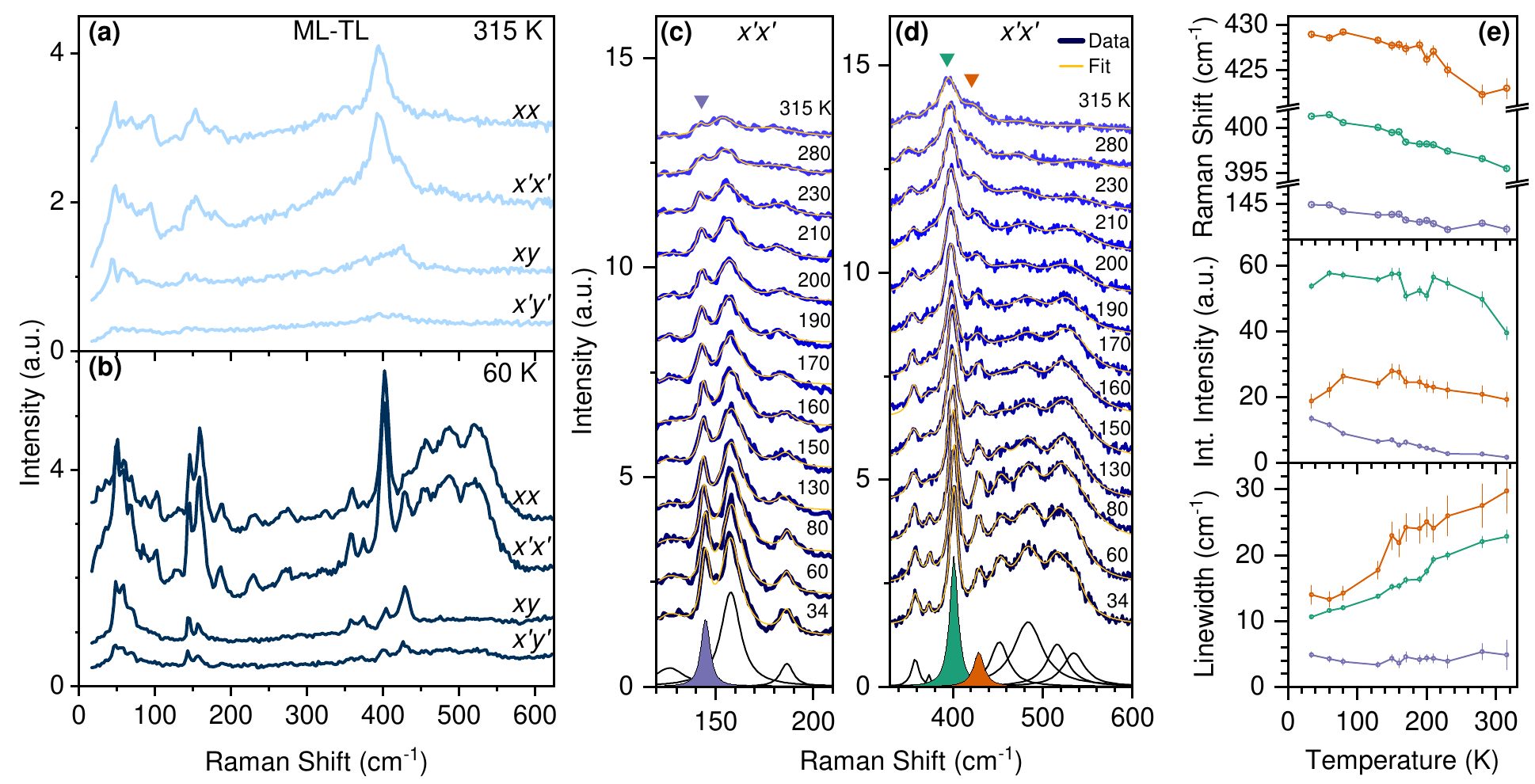}
    \caption{Polarization-resolved Raman spectra of the phonon sector of ML-TL at (a) 315~K and (b) 60 K, acquired in $xx$, $x'x'$, $xy$, and $x'y'$ scattering configuration, respectively. For clarity, the $xy$ and $x'x'$ spectra are offset in vertical direction by 0.5 a.u. and the $xx$ spectra by 2 a.u. (c) Temperature dependence of selected low-energy phonon modes. Spectra are offset in vertical direction by 1 a.u. Solid yellow lines are fits of the data using Voigt profiles. Individual profiles of the 34 K fit are shown as black lines at the bottom of the panel. The phonon mode around 140\,cm$^{-1}$ is highlighted by the triangle symbol and its fitted peak profile is shaded in purple. (d) Temperature dependence of higher-energy phonon modes, with modes around 400 and 425\,cm$^{-1}$ highlighted by green and orange triangle symbols, respectively, and the corresponding peak profiles shaded in the same colors. (e) Temperature dependence of the Raman shift, integrated intensity, and linewidth of the modes evolving around 140, 400, and 425\,cm$^{-1}$ with the same color coding as in panels (c) and (d). 
    }
    \label{fig:ML-TL_phonon}
\end{figure*}

Figure~\ref{fig:RoomT_Raman}(a) shows the Raman spectrum of oxygen-stoichiometric ML La$_2$NiO$_4$, which exhibits the orthorhombic space group $Bmab$ at room temperature. Notably, the symmetry of this material sensitively depends on the oxygen content \cite{Poirot1998}. In the present study, to enable the direct comparison between the Raman spectra of the different RP nickelates, the sample synthesis was optimized towards full oxygen stoichiometry of the compounds (for synthesis details see Materials and Methods and Supplementary Material). The spectrum in Fig.~\ref{fig:RoomT_Raman}(a) is characterized by three prominent low-energy phonon modes between 100-300 \,cm$^{-1}$ and an intense mode centered at 440 \,cm$^{-1}$, which involves buckling of the NiO$_6$ octahedra. Moreover, we observe a subtle mode around 670 cm$^{-1}$, which is indicative of the orthorhombic symmetry, whereas it is not present in the oxygen-excess tetragonal $I4/mmm$ phase of La$_2$NiO$_{4+\delta}$ \cite{RiazAM2025}. 

The BL compound La$_3$Ni$_2$O$_7$ crystallizes in the orthorhombic $Cmcm$ structure, although deviations from this symmetry have been reported, possibly reflecting differences in oxygen stoichiometry between samples studied in the literature \cite{Zhang1994JSSC}. The Raman spectrum of the BL phase is markedly distinct from that of the ML compound, with the most intense phonon mode centered at 580~cm$^{-1}$ and an additional prominent mode at 360~cm$^{-1}$ (Fig.~\ref{fig:RoomT_Raman}(b)). The former corresponds to an in-plane breathing distortion of the oxygen atoms within the NiO$_2$ planes, while the latter involves out-of-plane stretching of the apical oxygen atoms of the NiO$_6$ octahedra \cite{Zhang327HighP2025}. In addition, a broad and comparatively weak feature emerges around 1240~cm$^{-1}$, which was not reported in earlier Raman studies of the BL phase \cite{HeRSP2025}. 

For close-to-stoichiometric oxygen contents, the TL compound La$_4$Ni$_3$O$_{10}$ realizes the monoclinic space group $P2_1/a$, yielding a large number of phonon modes, particularly in the low-energy range between 60 and 200~cm$^{-1}$ (Fig.~\ref{fig:RoomT_Raman}(c)). The most prominent phonon mode appears near 400~cm$^{-1}$ \cite{LiSciBull2025,SutharVS2025}, clearly distinguishing the TL spectrum from those of the ML and BL compounds. This mode is associated with bond-stretching vibrations of basal oxygen atoms in the outer NiO$_2$ planes of the TL block \cite{SutharVS2025}. Notably, a broad continuum of spectral weight underlies the phonon peaks and extends up to approximately 910~cm$^{-1}$, where it terminates in a hump-like feature (green arrow in Fig.~\ref{fig:RoomT_Raman}(c)). Polarized Raman measurements have revealed a strong temperature dependence of this continuum, indicating gap formation associated with the intertwined CDW and SDW order in La$_4$Ni$_3$O$_{10}$ \cite{SutharVS2025}. As the hump feature defines the high-energy cutoff of this continuum, it is likely associated with an electronic excitation, such as a $2\Delta_{\mathrm{DW}}$ peak arising from short-range fluctuations of the DW that persist above the ordering temperature. Alternative interpretations cannot be fully excluded, including a multi-phonon scattering origin, and a recent unpolarized Raman study proposed a bimagnon origin \cite{zhangarXiv2025}.

Figure~\ref{fig:RoomT_Raman}(d) displays the Raman spectrum of the ML-TL polymorph. For oxygen-annealed single crystals with close-to-stoichiometric oxygen content the orthorhombic space group $Fmmm$ was determined by high-resolution single-crystal x-ray diffraction \cite{PuphalPRL2024}. In contrast, studies on as-grown crystals reported the orthorhombic space group $Cmmm$ \cite{ChenACS2024,WangInChem2024}. A key structural distinction between these two solutions is a doubling of the unit cell along the $c$ axis and the presence of octahedral tilts in the $Fmmm$ structure (Fig.~\ref{fig:Lattice}(d)). The Raman data shown here were acquired on an $Fmmm$ crystal from the same growth batch as those reported in Ref.~\onlinecite{PuphalPRL2024}. Remarkably, the ML-TL spectrum in Fig.~\ref{fig:RoomT_Raman}(d) closely resembles that of the pure TL phase (Fig.~\ref{fig:RoomT_Raman}(c)), including an intense phonon mode near 400~cm$^{-1}$ together with a broad electronic continuum and the hump-like feature around 910~cm$^{-1}$. Nevertheless, differences become apparent upon closer inspection of the low-energy phonon region, with several modes present in the TL Raman spectrum that are absent in the ML–TL phase, as indicated by the vertical dashed lines in Figs.~\ref{fig:RoomT_Raman}(c) and (d).

Overall, Fig.~\ref{fig:RoomT_Raman} establishes that Raman spectroscopy provides clear, phase-specific fingerprints across the RP nickelate family, that are useful for phase identification in the context of crystal growth and sample characterization, which has proven challenging in this class of materials \cite{Zhou2025a}. The close resemblance between the TL and ML–TL spectra naturally reflects the presence of TL building blocks in the polymorph, while obvious signatures associated with the ML block appear elusive in the ML–TL spectrum. To further disentangle how both the ML and TL building blocks contribute to the lattice dynamics of the ML–TL phase, we next analyze its individual phonon modes.

\subsection{Phonons}

\begin{figure*}[tb]
    \centering
    \includegraphics[width=\linewidth]{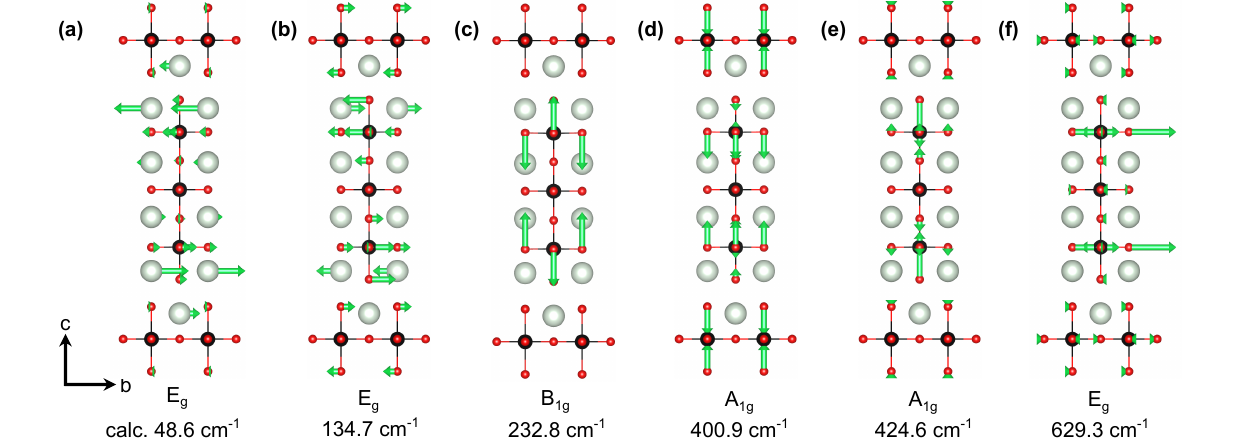}
    \caption{(a-f) Atomic displacement patterns of selected phonon modes of ML-TL for the tetragonal $P4/mmm$ unit cell. The $E_g$, $B_{1g}$, and $A_{1g}$ mode symmetries are indicated along with the computed phonon frequencies. Green arrows display the directions and amplitudes of the atomic vibrations. For clarity, only displacements with large amplitudes are shown for each mode. The $b$ and $c$ directions of the $P4/mmm$ unit cell are also indicated.
    }
    \label{fig:ML-TL_theory}
\end{figure*}

The $Fmmm$ crystal structure of ML-TL ($D_{2h}$ point group) gives rise to 131 vibrational modes at the $\Gamma$ point, out of which 63 are Raman active (19 A$_g$ + 10 B$_{1g}$ + 17 B$_{2g}$ + 17 B$_{3g}$) and 68 are infrared active. The corresponding Raman tensors are presented in the Supplemental Material. Figures~\ref{fig:ML-TL_phonon}(a) and (b) show Raman spectra of the ML–TL compound measured in four polarization configurations ($xx$, $x'x'$, $xy$, and $x'y'$) at 315 and 60 K, respectively. According to the polarization selection rules for our employed backscattering geometry, only A$_g$ and B$_{1g}$ modes are expected to be observable, with the B$_{1g}$ modes suppressed in the $xx$ and $x'y'$ configurations. Consistent with this expectation, phonon modes such as those near 355 and 425~cm$^{-1}$ exhibit significantly enhanced intensity in the $x'x'$ and $xy$ channels compared to the $xx$ and $x'y'$ channels, respectively. Residual intensity observed in the nominally forbidden channels may arise from polarization leakage or from microscopic $a/b$ twinning within the laser spot. By contrast, the $x'x'$ configuration allows contributions from both A$_g$ and B$_{1g}$ modes, and the corresponding spectrum in Fig.~\ref{fig:ML-TL_phonon}(b) exhibits the highest number of modes among the four channels, with at least 23 individual peaks discernible.  

The temperature dependence of several representative phonon modes between 120 and 600~cm$^{-1}$ in the $x'x'$ channel is displayed in Figs.~\ref{fig:ML-TL_phonon}(c) and (d). Upon cooling, both the low-energy modes in Fig.~\ref{fig:ML-TL_phonon}(c) and the modes near 400~cm$^{-1}$ in Fig.~\ref{fig:ML-TL_phonon}(d) sharpen and gain intensity. In addition, a prominent cluster of modes emerges between 430 and 580~cm$^{-1}$, with continuously increasing intensity down to the lowest measured temperatures (Fig.~\ref{fig:ML-TL_phonon}(c)).

The detailed temperature evolution of the phonon energy, integrated intensity, and linewidth for representative modes near 140~cm$^{-1}$ (purple triangle in Fig.~\ref{fig:ML-TL_phonon}(c)), 400~cm$^{-1}$ (green triangle in Fig.~\ref{fig:ML-TL_phonon}(d)), and 425~cm$^{-1}$ (orange triangle in Fig.~\ref{fig:ML-TL_phonon}(d)) is summarized in Fig.~\ref{fig:ML-TL_phonon}(e). These parameters were extracted from Voigt-profile fits to the individual phonon peaks. Overall, the temperature evolution is  continuous within the experimental error, without clear anomalies at specific temperatures that would indicate a structural or electronic phase transition. This behavior contrasts with the Raman spectra of the TL compound, where additional phonon modes appear below the DW transition around 140 K and subtle renormalizations of phonon self-energy parameters have been reported \cite{SutharVS2025}.

To identify the hallmark phonons of the ML-TL polymorph, we next perform DFPT calculations in combination with the PHONOPY package (see Materials and Methods). Due to the excessive size of the $Fmmm$ unit cell of ML–TL (Fig.~\ref{fig:Lattice}(d)), we instead adopt the smaller $P4/mmm$ unit cell with half the $c$-axis lattice parameter and without octahedral tilts for the phonon calculations. Note that pressure transforms the $Fmmm$ structure of ML–TL into $P4/mmm$ \cite{PuphalPRL2024}, indicating a close structural relationship between the two structures and their hallmark phonons. The $P4/mmm$ unit cell ($D_{4h}$ point group) gives rise to 48 phonons at the $\Gamma$ point, out of which 18 are Raman active (8 A$_{1g}$ + 9 E$_g$ + 1 B$_{1g}$) and 30 are infrared active. A table with the calculated phonon frequencies is provided in the Supplementary Material. 

Figures~\ref{fig:ML-TL_theory}(a–f) show the atomic displacement patterns of representative phonon modes. The lowest computed mode at 48.6~cm$^{-1}$ has $E_g$ symmetry and involves in-plane oxygen displacements in the TL block, together with subtle displacements of apical oxygen atoms in the ML block. In addition, La atoms from both the ML and TL blocks participate in the distortion, with the largest displacements occurring on the La atoms in the outermost TL layer (Fig.~\ref{fig:ML-TL_theory}(a)). A phonon at a comparable energy is observed experimentally (Figs.~\ref{fig:ML-TL_phonon}(a) and (b)), suggesting a correspondence to the symmetry-related mode of the $Fmmm$ structure. A second $E_g$ mode, calculated at 134.7~cm$^{-1}$, exhibits similar in-plane oxygen motion, but without participation of the ML La atoms (Fig.~\ref{fig:ML-TL_theory}(b)). Experimentally, a strong phonon appears near 134~cm$^{-1}$ in the $xx$ and $x'x'$ channels, and a mode at similar energy is also present in the pure TL compound \cite{SutharVS2025}, indicating that several modes are intrinsic to the TL building block and recur in both systems.

\begin{figure*}[tb]
    \centering
    \includegraphics[width=\linewidth]{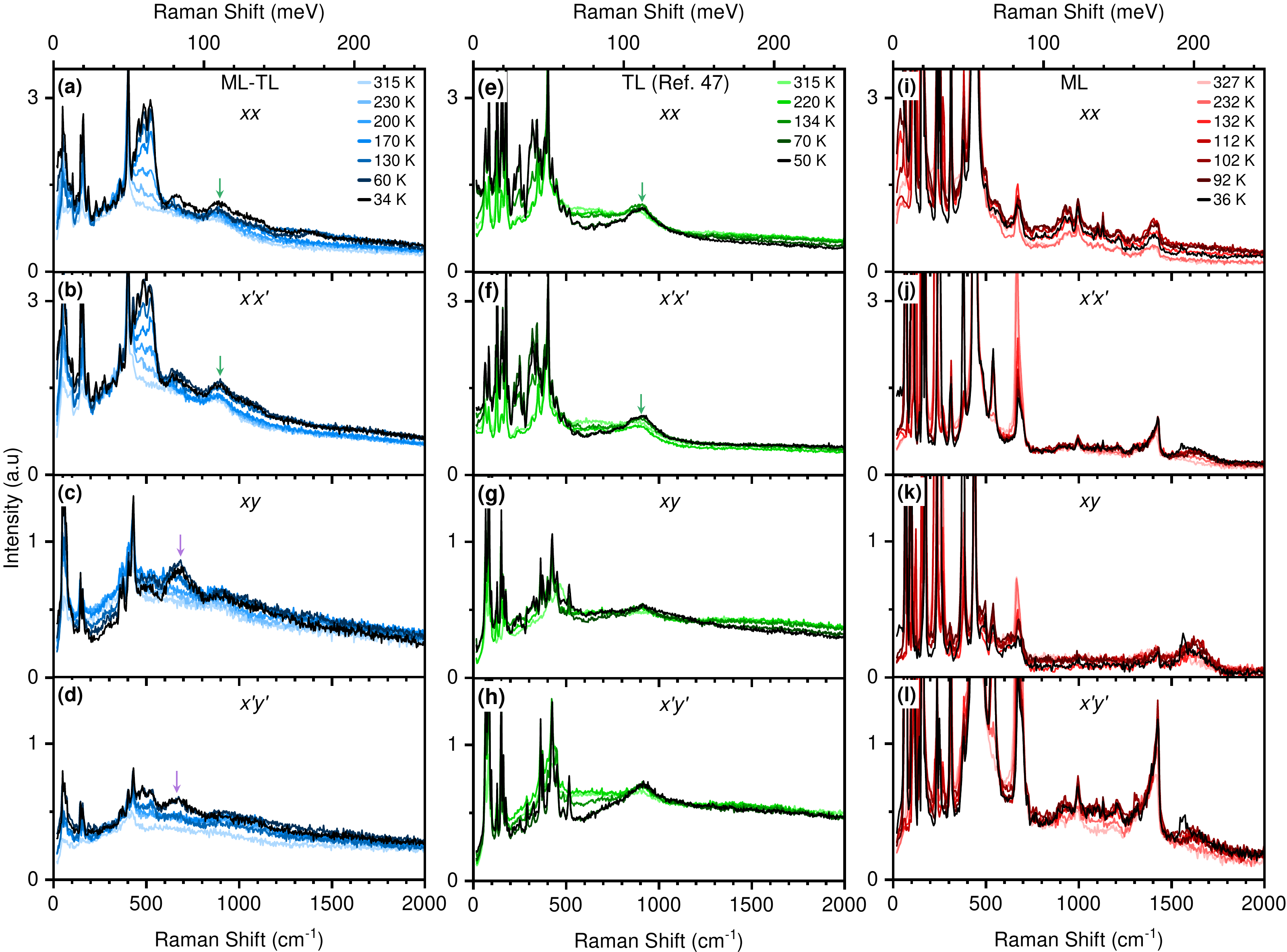}
    \caption{Electronic Raman scattering across a wide energy range of (a-d) ML-TL, (e-h) TL, and (i-l) ML phases. Spectra were acquired at different temperatures (light to dark shades) and in $xx$, $x'x'$, $xy$, and $x'y'$ scattering configurations (top to bottom panel), respectively. Green arrows mark a feature around 900 cm$^{-1}$ that occurs most prominently in the $xx$ and $x'x'$ channels of the ML-TL and TL phases. Purple arrows highlight a feature occurring most pronounced in the $xy$ and $x'y'$ channels of ML-TL. The TL spectra in panel (e-h) are adopted from Ref.~\onlinecite{SutharVS2025}.
   }
    \label{fig:LowT_Raman}
    \end{figure*}

The only $B_{1g}$ mode in the $P4/mmm$ structure is computed at 232.8~cm$^{-1}$ and is characterized by out-of-plane displacements of in-plane oxygen atoms in the outer TL layers, with nearest neighbors moving anti-parallel within the plane (Fig.~\ref{fig:ML-TL_theory}(c)). 
In the vicinity of 400~cm$^{-1}$, the calculations yield several A$_{1g}$ modes, including a mode at 400.9~cm$^{-1}$ that involves axial displacements of in-plane oxygen atoms against the NiO$_2$ planes in both the ML block and the outer TL layers (Fig.~\ref{fig:ML-TL_theory}(d)). Experimentally, a dominant phonon peak is observed near 400~cm$^{-1}$ in both the ML–TL and pure TL compounds. In the latter case, however, this mode has been identified as an A$_g$ vibration dominated by in-plane bond-stretching displacements of basal oxygen atoms in the outer NiO$_2$ planes \cite{SutharVS2025}. These distinct displacement patterns indicate that the similarity in frequency to the ML–TL mode involving ML-block displacements might be coincidental. An additional computed A$_{1g}$ mode in the $P4/mmm$ structure at 424.6~cm$^{-1}$ involves out-of-plane displacements of apical oxygen atoms in the outer TL layer, while distortions in the ML block are small (Fig.~\ref{fig:ML-TL_theory}(e)).

The highest-energy computed mode at 629.3~cm$^{-1}$ exhibits $E_g$ symmetry and is dominated by in-plane horizontal displacements of oxygen and Ni atoms in the outer TL layer (Fig.~\ref{fig:ML-TL_theory}(f)). Whether this type of displacement pattern is representative of all modes within the broad high-energy phonon cluster between 430 and 580~cm$^{-1}$ remains to be assessed by models that incorporate full structural details beyond those captured by the $P4/mmm$ calculation.

Taken together, our phonon analysis indicates that the ML unit in the ML–TL polymorph is not merely a passive structural spacer but contributes actively and in a mode-selective manner to the lattice dynamics. In particular, several experimentally relevant phonon modes, including those near 48, 134, and 400~cm$^{-1}$, likely involve atomic displacements within the ML block.

\subsection{Electronic Raman response}

As a next step, we examine the polarization and temperature dependence of the electronic Raman response over an extended energy range up to 2000~cm$^{-1}$. For the ML–TL polymorph, the hump feature at 910~cm$^{-1}$ defines an upper energy scale beyond which the spectra in all four polarization channels are largely featureless and exhibit only a weak temperature dependence (Figs.~\ref{fig:LowT_Raman}(a–d)). This feature is most pronounced in the parallel polarization channels (green arrows in Figs.~\ref{fig:LowT_Raman}(a,b)) and is strongly suppressed in the crossed channels (Figs.~\ref{fig:LowT_Raman}(c,d)). These characteristics closely resemble those reported for the corresponding 910~cm$^{-1}$ feature in the TL compound (see Ref.~\onlinecite{SutharVS2025} and green arrows in Figs.~\ref{fig:LowT_Raman}(e,f)), where it has been interpreted either as the $2\Delta_{DW}$ peak associated with an intertwined DW \cite{SutharVS2025} or as a bimagnon excitation \cite{zhangarXiv2025}.

Remarkably, the ML–TL phase exhibits an additional distinct peak-like feature near 680~cm$^{-1}$ at low temperatures, with a linewidth comparable to that of the 910~cm$^{-1}$ hump. This feature is most prominent in the $xy$ polarization channel (purple arrow in Fig.~\ref{fig:LowT_Raman}(c)), although weaker signatures are also visible in the other channels. No analogous feature is observed in the TL compound (Figs.~\ref{fig:LowT_Raman}(e–h)). Instead, the TL spectra show a pronounced suppression of spectral weight in this energy range upon cooling, extending from low energies up to the 910~cm$^{-1}$ feature (Figs.~\ref{fig:LowT_Raman}(e,f)), which has been associated with the incoherent opening of the DW gap \cite{SutharVS2025}. In ML–TL, a partial suppression of low-energy spectral weight is also discernible in the $xy$ channel between 100 and 350~cm$^{-1}$. However, because this effect is essentially absent in the other polarization channels (Figs.~\ref{fig:LowT_Raman}(a,b,d)), a direct analogy to the TL behavior is not straightforward.

These observations raise the question of whether the presence of the ML block can directly account for the additional 680~cm$^{-1}$ feature unique to the ML–TL phase. To address this, we performed reference measurements on the pure ML compound (Figs.~\ref{fig:LowT_Raman}(i–l)). While the ML spectra indeed exhibit several high-energy features, these are generally much sharper than those observed in the ML–TL and TL compounds between 500 to 910~cm$^{-1}$. This is also the case for a prominent feature near 680~cm$^{-1}$ in the ML material, which is most intense in the $x'x'$ and $x'y'$ polarization channels and can be assigned to a phonon mode. Moreover, in contrast to the ML–TL spectra, this mode in the ML compound shows a significant reduction in spectral weight upon cooling from 327 to 36~K (Fig.~\ref{fig:LowT_Raman}(j)), consistent with a phononic origin and with its relation to the orthorhombic-to-tetragonal structural phase transition of the ML compound at low temperatures \cite{RiazAM2025}. Several of the higher-energy features in ML are multi-phonon peaks and the broad feature around 1700~cm$^{-1}$ likely corresponds to a bimagnon \cite{Sugai1998}. 

Nevertheless, an electronic Raman feature reminiscent of the 680~cm$^{-1}$ peak in ML-TL has been reported in the BL compound below 150~K, where it was attributed to SDW formation \cite{HeRSP2025}. Notably, in the BL compound the characteristic 650~cm$^{-1}$ feature emerges exclusively in the $xy$ channel, whereas in our ML–TL spectra it persists in multiple polarization channels, albeit with reduced intensity. Furthermore, BL exhibits a second SDW-related feature appearing only in the $x'y'$ channel near 370~cm$^{-1}$. In ML–TL, spectral weight gain at low temperatures is also observed in the $x'y'$ channel around 500~cm$^{-1}$, which however might be associated with increasing intensity of closely spaced phonon modes in this energy range.  

The temperature evolution of the summed integrated intensity of the modes within the cluster from 430 to 600~cm$^{-1}$ is continuous over the full temperature range investigated, consistent with that of phonons in ML-TL (see Supplementary Material). By contrast, the integrated intensity of the 680~cm$^{-1}$ feature exhibits a qualitatively distinct temperature dependence, with an onset at approximately 170~K. These contrasting behaviors support our assignment of the modes within the cluster to a phononic origin, whereas the 680~cm$^{-1}$ feature is attributed to electronic Raman scattering associated with DW formation. More broadly, since none of the low-temperature emergent features can be directly traced to an origin in the isolated ML block, the electronic Raman response of the ML–TL polymorph cannot be understood as a simple superposition of its ML and TL constituents, but instead reflects a distinct electronic structure intrinsic to the ML–TL phase.

\section{Discussion}

Our comparative Raman study reveals a nuanced interplay between lattice dynamics and electronic excitations in the ML–TL polymorph, highlighting both similarities to and important departures from the behavior of the pure TL and BL compounds. At room temperature, the phononic Raman response of the ML–TL phase closely resembles that of the pure TL compound. However, a combined analysis of phonon frequencies and first-principles displacement patterns demonstrates that atomic motions within the ML block contribute significantly to a subset of phonon modes. While the TL block remains a dominant contributor to the lattice dynamics, the ML block is clearly not a passive structural spacer. Instead, it hybridizes with TL-derived vibrations and actively reshapes the phonon spectrum of the ML–TL polymorph.

Beyond similarities and subtle differences to the TL spectra at room temperature, unique characteristics of the phononic Raman response of the ML–TL phase emerge upon cooling. In particular, a pronounced cluster of modes emerges between 430 and 600~cm$^{-1}$, most prominently in the $xx$ and $x'x'$ polarization channels. The continuous temperature evolution of this spectral weight, together with its polarization dependence, is characteristic of phononic rather than electronic excitations. This interpretation is further supported by an analysis of the temperature dependence of the integrated intensity in the 120–210~cm$^{-1}$ and 430–600~cm$^{-1}$ ranges (see discussion in the Supplementary Material).

In stark contrast to the continuous phononic Raman response as a function of temperature, the electronic Raman scattering shows qualitatively new features at low temperatures. Specifically, a broad feature develops near 680 cm$^{-1}$ that has no direct analogue in the TL compound and also differs from the polarization selection rules of the SDW feature reported for the BL phase \cite{HeRSP2025}. The distinctive symmetry-dependent redistribution of spectral weight at low temperatures indicates that a simple transfer of the BL SDW scenario to the ML–TL phase is insufficient. Furthermore, the vanishing of spectral weight associated with the DW gap opening in the pure TL compound \cite{SutharVS2025} is absent or superposed by the emergence of the 680 cm$^{-1}$ feature in ML-TL. Nevertheless, the feature near 900~cm$^{-1}$ attributed to short-range CDW fluctuations in pure TL \cite{SutharVS2025} occurs closely similar in ML-TL across the different polarization channels, indicating comparable CDW correlations in the TL block of ML-TL.  

These findings suggest that the ML–TL polymorph does not simply inherit the DW physics of the TL or the BL compound, but instead realizes an at least partially distinct correlated electronic state. This conclusion is consistent with recent theoretical work predicting that the periodic insertion of ML blocks leads to a more complex electronic structure than in the pure TL phase, including additional Fermi surface sheets and enhanced correlations \cite{LechermannPRM2024,LaBollitaPRB2024}.  The ML blocks therefore not only modify the lattice dynamics but also qualitatively alter the low-energy electronic excitations probed by Raman scattering.

\section{Conclusion}

In summary, our study compared the unique Raman spectroscopic fingerprints of the RP family members ML La$_2$NiO$_4$, BL La$_3$Ni$_2$O$_7$, TL La$_4$Ni$_3$O$_{10}$, and the polymorph ML-TL La$_3$Ni$_2$O$_7$, providing clear benchmarks for phase identification and characterization studies. A central result is that, although the room-temperature Raman spectra of ML-TL and TL appear closely similar, pronounced differences emerge at low temperatures, particularly affecting the polarization-resolved electronic Raman response. We attribute these contrasts to subtle but important differences in the underlying electronic structures of the two materials, driven by self-doping and quantum-confinement effects from the ML unit in the ML-TL crystal structure. Understanding how these differences relate to superconductivity, in particular to the possibly reduced $T_c$ of ML-TL compared to TL and BL, will be an important direction for future work.

 \begin{acknowledgements}
We thank A.~Greco, M.~J.~Graf von Westarp, and A.~von Ungern-Sternberg Schwark for insightful discussions.
\end{acknowledgements}

\bibliography{bibliography}
\end{document}